\documentclass[aps,pra,10pt,notitlepage,twocolumn]{revtex4-1}
\usepackage[dvips]{graphicx}
\usepackage{amssymb,amsfonts}
\usepackage{amsmath}

\begin{document}

\author{J. Chwede\'nczuk$^1$, F. Piazza$^2$ and A. Smerzi$^{3}$}
\affiliation{
  $^1$Faculty of Physics, University of Warsaw, ul.\ Ho\.{z}a 69, PL--00--681 Warszawa, Poland\\ 
  $^2$Physik Department, Technische Universit\"at M\"unchen, 85747
  Garching, Germany\\
  $^{3}$QSTAR, INO-CNR and LENS,  Largo Enrico Fermi 2, 50125 Firenze, Italy 
  \\
}


\title{A Multi-path Interferometer with Ultracold Atoms Trapped in an Optical Lattice}

\begin{abstract}
We study an ultra-cold gas of $N$ bosons trapped in a one dimensional $M$-site
optical lattice perturbed by a spatially dependent potential $g\cdot x^j$, where the unknown
coupling strength $g$ is to be estimated. We find that the measurement
uncertainty is bounded by $\Delta g\propto\frac1{N (M^j-1)}$. For a typical case of a linear potential, the sensitivity improves as $M^{-1}$, 
which is a result of multiple interferences
between the sites -- an advantage of multi-path interferometers over the two-mode setups. 
Next, we calculate the estimation sensitivity for a
specific measurement where, after the action of the potential, the particles are
released from the lattice and form an interference pattern.  
If the parameter is estimated by
a least-square fit of the average density to the interference
pattern, the sensitivity still scales like $M^{-1}$ for
linear potentials
and can
be further improved by preparing a properly correlated initial state
in the lattice. 

\end{abstract}
\maketitle

\section{Introduction}

As scientists gain insight into the ``nano'' scale in both the space and time domain, it becomes increasingly relevant to develop ultra-precise measurement devices.
Very prominent among those are  interferometers, where the
parameter to be measured is mapped into a phase difference $\theta$ (or optical
path length difference) between two
(or more) arms of the device. By recombining the outputs of each arm,
an interference figure appears, which contains information about the
parameter to be estimated.

The key quantity which describes the performance of an interferometer is its precision $\Delta\theta$, 
often referred to as the sensitivity. In a typical interferometer, where the signal which is used
to determine $\theta$ comes from a collection of $N$ independent particles, the sensitivity scales at the shot-noise level, 
i.e. $\Delta\theta\propto\frac1{\sqrt m}\frac 1{\sqrt N}$, where $m$
denotes the number of independent repetitions of the measurement. 
This expression tells that the sensitivity can be improved by either
increasing the number of trials $m$ or the number of probe particles
$N$. A third way of decreasing $\Delta\theta$ is by replacing the uncorrelated ensemble of atoms with a usefully entangled state. 
Non-classical correlations can be employed to beat the Shot-Noise Limit (SNL) and
ultimately reach the Heisenberg Level (HL), where
$\Delta\theta\propto\frac1{\sqrt m}\frac 1{N}$, which for large $N$
can be a significant improvement over the SNL \cite{giovanetti}. 

A paradigmatic example of an interferometer which benefits from the use
of entangled states is a Mach-Zehnder Interferometer (MZI), 
where the signal coming from two arms at the input is first recombined/split using a beam-
splitter, then the phase difference is
imprinted between the two arms, and finally another beam-splitter
maps the phase difference into the intensities in the output ports.
When the MZI is fed with an entangled spin-squeezed state and the
phase $\theta$ is estimated from the measurement of the population
imbalance between these ports, 
the Sub Shot-Noise (SSN) sensitivity $\Delta\theta<\frac1{\sqrt m}\frac1{\sqrt N}$ can be acheived. 

While originally implemented with photons, in recent years interferometers fed by
atoms have attracted a lot of interest \cite{cronin}, and will be the object of the present work. These interferometers
proved promising for the precise determination of electromagnetic 
\cite{ketterle_casimir,vuletic_casimir,cornell} and gravitational \cite{kasevich,fattori,hinds} interactions. 
Moreover, their performance can be improved with the recently created non-classical states of of matter \cite{cold1,cold2,cold3,gross,esteve,riedel,maussang,smerzi}.
In \cite{esteve,riedel,maussang} a spin-squeezed state was generated by reducing fluctuations between the sites of a double-well
potential, while in \cite{gross,app,chen,smerzi} internal degrees of freedom were employed.
In the experiments mentioned above, a two-path (or two mode) usefully
entangled quantum state was prepared.
In this work, we consider instead a multi-path atom interferometer, involving more than two internal/external degrees of freedom.

The idea that the growing number of modes could improve the performance of an interferometer has been first discussed in the context of optical devices \cite{zernicke,paris}. 
They involve the multi-mode version of a beam-splitters, and allow in general to measure a vector phase
\cite{sanders}. In the particular case where the phase difference is
the same between each pair of ``neighboring'' paths, different
methods have been used to derive bounds for the phase sensitivity. In
\cite{paris}, for an input coherent state, a momentum formula
approximation allowed to predict a sensitivity
$\Delta\theta\propto M^{-1}$, where $M$ is the number of ports. This is a
result of a multiple interference between different paths, and does not
violate the HL \cite{inoue}. The gain is also robust against losses \cite{paris}, differently from the improvement due to entanglement present in the input state. 
The scaling of the sensitivity with the number of modes and particles has also been heuristically derived for
the specific case where coherent or number states are used in input \cite{vourdas}.
Recently, the sensitivity for rotation measurements using a ring interferometer with ultracold atoms inside a lattice
has been studied in \cite{dunningham}. Numerical determination of the
ground state allowed to calculate the Quantum Fisher Information (QFI) for
specific states, which, as we will discuss in the following, provides the ultimate sensitivity bound,
independently of the measurement and estimation strategy employed. 
On the experimental side, a multimode atom interferometer was implemented
in \cite{petrovic}, where a single Bose-Einstein Condensate was coherently split among five Zeeman sublevels using a radio frequency pulse. 
The relative phase between the atoms occupying different substates was
imprinted with an external magnetic field. Finally, the cloud was recombined with another pulse and an increased interferometric signal
-- with respect to the two-mode case -- was observed. 
A multi-mode nonclassical state of ultracold bosons inside an optical lattice has
also recently been created \cite{oberthaler_lattice}

In this manuscript, we consider a multi-path interferometer realized
with a dilute gas of ultracold $N$ bosons trapped inside a one-dimensional
optical lattice \cite{rmp_zw,bloch,bloch2}. Each of the $M$ lattice sites plays the role of a possible
interferometric path, whose optical length can be modified by
adding a spatially dependent force which would be then the object of
our precision measurement. More precisely, for a 
potential $V(x)=gx^j$, we derive the sensitivity for the estimation of
the parameter $g$. First, using the Quantum Fisher Information (QFI), we derive the best possible
precision to be $\Delta g\propto\frac1{\sqrt{m} N (M^j-1)}$, independently of the quantum state, output
measurement and estimation strategy. Then, choosing a specific class
of states which are symmetric with respect to exchange of lattice
sites
, we obtain the sensitivity $\Delta g\propto\frac1{\sqrt{m
  f(N)}M^j}$ for $M\gg 1$. Here, $f(N)$ is a two-mode analog of the enanglement measure, which can vary from $N$ for 
the superfluid state of the lattice up to $N^2$ for a strongly entangled state.
Note that when the potential $V(x)$ acts on the system, one can naturally define a phase $\theta=gx_0^jt/\hbar$, which is related to the coupling constant, the duration
of the interaction $t$ and the lattice spacing $x_0$. In spirit of the two-mode interferometry, one can thus speak of the phase estimation, rather then determination of the coupling
constant $g$. Throughout this work, we use both $\Delta g$ and $\Delta\theta$, and keep in mind they are simply related by $\Delta g=(\hbar/t x_0^j)\Delta\theta$.

We finally consider a specific measurement and estimation strategy
which could be applied in most of the current ultracold atom
experimental setups: after undergoing the action of the force, the atoms are
released from the lattice, form an interference pattern, and the
parameter is extracted by a least-squares fit of the average density.
Using rigorous results from the estimation theory \cite{chwed_njp}, we
derive the sensitivity for this interferometric protocol, and show
that it scales with $M$ in a same manner as in the case of the optimal measurement. 
The use of the spatial interference pattern, obtained through the
expansion of the cloud, substitutes the beam splitter. The in-situ implementation of the latter in an
optical lattice, in analogy to the double-well implementation,  seems a much more demanding task.



The manuscript is organized as follows. In Section \ref{formula} we formulate the framework for analyzing the performance of the multi-mode interferometers.
In Section \ref{abst}, we employ the notion of the QFI to provide some ultimate bounds for the multi-mode interferometry. First, in Section \ref{abst_ult}
we identify the states, which allow to reach the best possible scaling of $\Delta\theta$ with the number of sites $M$ and linear potentials $V(x)=g\,x$. 
Then in Section \ref{abst_sym}, using the QFI, we show how 
$\Delta\theta$ scales with $M$ for states which are symmetric with respect to interchange of any two sites. In Section \ref{nonlin} we generalize these results for non-linear potentials
$V(x)=g\,x^j$.
Finally, in Section \ref{fit} we show how the scaling with the number of sites can be achieved
for a particular detection scheme and identify the states usefully entangled for the multi-mode interferometry.
Some details of analytical calculations are presented in the Appendices.

\section{Formulation of the model}
\label{formula}


We begin the analysis of multi-mode interferometry by introducing the quantum state $|\psi\rangle$ of $N$ bosons distributed among $M$ modes. 
The most general expression for $|\psi\rangle$ is
\begin{subequations}\label{gen1}
  \begin{eqnarray}
    &&|\psi\rangle=\sum_{n_1=0}^N\sum_{n_2=0}^{N-n_1}\ldots\!\!\!\!\!\sum_{n_{M-1}=0}^{N-n_1-\ldots-n_{M-2}}\!\!\!\!C_{n_1\ldots n_{M-1}}\\
    &&|n_1,n_2,\ldots ,n_{M-1},N-n_1-\ldots -n_{M-1}\rangle,
  \end{eqnarray}
\end{subequations}
where the sum runs over occupations of $M-1$ wells. Since the total number of atoms is set to be $N$, the population of the $M$-th mode is determined by other $M-1$ occupancies. 
The coefficient $C_{n_1\ldots n_{M-1}}$ is the probability amplitude for having $n_1\ldots n_{M-1}$ atoms in the $M-1$ modes and $N-(n_1+\ldots n_{M-1})$ in the last one.

For future purposes we introduce a shortened notation, where the set of $M-1$ indeces $(n_1\ldots n_{M-1})$ is represented by a vector $\vec n$, and the state (\ref{gen1}) is written as
\begin{equation}\label{state}
  |\psi\rangle=\sum_{\vec n}C_{\vec n}|\vec n\rangle.
\end{equation}
The complete description of quantum properties of the system includes the $M$-mode field operator which reads
\begin{equation}\label{field}
  \hat\Psi(x)=\sum_{k=1}^M\psi_k(x)\hat a_k.
\end{equation}
Here, $\hat a_k$ is a bosonic operator which annihilates an atom occupying the mode-function $\psi_k(x)$ localized in the $k$-th well.

Let us for now restrict ourselves to the case where a constant force,
corresponding to the linear potential $V(x)=g\,x$, is added to the
lattice potential (the more general case will be consider in Section \ref{nonlin}). Here $g$ is the
coupling constant - the parameter which we want to estimate. 

The Hamiltonian corresponding to the added potential, in the second quantization, reads
\begin{equation}\label{ham}
  \hat H=\epsilon\sum_{k=1}^Mk\cdot\hat n_k,
\end{equation}
where $\hat n_k=\hat a^\dagger_k\hat a_k$ is the on-site atom number
operator and $\epsilon=g\cdot x_0$, where $x_0$ is the lattice
spacing. 

The external force interacts with the trapped gas for a time $t$, which transforms the field operator (\ref{field}) as follows
\begin{equation}\label{ev_op}
  \hat\Psi(x|\theta)\equiv e^{i\theta\hat h}\hat\Psi(x)e^{-i\theta\hat h}=\sum_{k=1}^M\psi_k(x)\hat a_ke^{-ik\theta},
\end{equation}
where $\theta=\epsilon t/\hbar$ and $\hat h$ is the generator of the phase-shift transformation
\begin{equation}\label{gen}
  \hat h=\sum_{k=1}^Mk\cdot\hat n_k.
\end{equation}
Equation (\ref{ev_op}) shows that the perturbation imprints a phase $\theta$ between two neighboring sites, while 
the phase between the two most distant wells is $(M-1)\theta$. 

The interferometric sequence consists thus of a simple phase shift
acting on the input state (\ref{state}) and could be implemented as follows. First, is the preparation of
the input state (\ref{state}) in an unperturbed
optical lattice ($g=0$). Then the perturbing
potential is suddenly turned on on a time scale shorter than the lattice hopping
time, but slow enough to avoid creating too many excitations. In this way, the state does not follow the change adiabatically, and the evolution
operator is  $\exp(-i\hat h\theta)$.
The goal of this work is to derive the sensitivity of the estimation
of $\theta$ or, equivalently, of
the coupling constant $\Delta g=(\hbar/t x_0)\Delta\theta$.

Given the interferometric operation defined by
Eqs.~(\ref{ev_op}),(\ref{gen}), the sensitivity $\Delta\theta$ still
depends on: a) the choice of the \emph{input state} $|\psi\rangle$, b) the
choice of the \emph{measurement} to be performed on the final state
$\exp(-i \hat h \theta )|\psi\rangle$, and c) the choice of the
\emph{estimator}, that is, how to infer the value of the parameter given the
measurement outputs. In Section \ref{abst}, we will derive the
\emph{optimal sensitivity}, that is, assuming that both the measurement and the estimator are
the best achievable. This can be done using the notion of the QFI. 
The latter allows to calculate the optimal
sensitivity for a given input state. In Sections \ref{abst_ult},~\ref{nonlin}, by
further optimizing the QFI over all possible input states, we derive
the ultimate scaling of the sensitivity. In Section \ref{abst_sym}
instead, we restrict to a particular family of input states which are
symmetric with respect to the exchange of two lattice sites. Finally,
in Section \ref{fit}, we derive the sensitivity for a particular
measurement and estimator, using the familty of input states used in
Section \ref{abst_sym}.

\section{Sensitivity for the optimal measurement}
\label{abst}

\subsection{Ultimate scaling}
\label{abst_ult}

As anticipated, using the notion of the QFI, denoted here as $F_Q$, we
can calculate the sensitivity for the best possible measurement and
estimator. 
Once we know $F_Q$, we can indeed employ the Cramer-Rao Lower Bound (CRLB)\cite{nota_qfi}:
\begin{equation}\label{crlb}
  \Delta^2\theta\geq\frac1m\frac1{F_Q},
\end{equation} 
where $m$ is the number of independent repetitions of the measurement
used to estimate $\theta$. The right side of Eq.~(\ref{crlb}) is the
sensitivity optimized over all possible measurements and  estimators.

The QFI yet depends on the interferometric operation and
the input state.
For pure states, the value of $F_Q$ is given by four times
the variance of the generator of the interferometric operation, in our
case the operator $\hat h$
generating the phase shift, calculated with the input state of the system and reads
\begin{equation}\label{qfi}
  F_Q=4\left[\langle\psi|\hat h^2|\psi\rangle-\langle\psi|\hat h|\psi\rangle^2\right]\equiv4\Delta^2\hat h.
\end{equation}
The QFI can be maximized with respect to $|\psi\rangle$ and the upper bound for $F_Q$ is provided by a simple relation
\begin{equation}\label{var}
  \Delta^2\hat h\leq\frac14(\lambda_{\rm max}-\lambda_{\rm min})^2,
\end{equation}
where $\lambda_{\rm min/max}$ are the smallest/largest eigenvalues of the operator $\hat h$. The two eigenvectors of $\hat h$ with extreme eigenvalues are
\begin{subequations}\label{minmax}
  \begin{eqnarray}
    &&|\psi_{\rm min}\rangle=|N,0,\ldots,0\rangle\rightarrow\hat h|\psi_{\rm min}\rangle=N|\psi_{\rm min}\rangle,\phantom{abcdefs}\\
    &&|\psi_{\rm max}\rangle=|0,0,\ldots,N\rangle\rightarrow\hat h|\psi_{\rm max}\rangle=MN|\psi_{\rm max}\rangle.
  \end{eqnarray}
\end{subequations}
By combining Equations (\ref{var}) and (\ref{minmax}) we get $F_Q\leq N^2(M-1)^2$. The inequality is saturated for the NOON state
$\frac1{\sqrt2}\left(|\psi_{\rm min}\rangle+|\psi_{\rm max}\rangle\right)$ for which, according to Eq.~(\ref{crlb}), the sensitivity is bounded by
\begin{equation}
  \Delta^2\theta\geq\frac1m\frac1{N^2}\frac1{(M-1)^2}.\label{opt}
\end{equation} 
This expression is the first important result: when the perturbation
is a linear potential (constant force), the best sensitivity reachable scales like
$M^{-1}$ -- a kind of ``Heisenberg scaling'' with the number of wells. In section \ref{nonlin}, we will derive the equivalent bound
for nonlinear potentials.

As anticipated in the introduction, this scaling is a purely geometrical -- and thus not quantum -- effect. While the Heisenberg scaling with the number of atoms $N$ originates from
a macroscopic quantum superposition present in the NOON state, the
scaling with $M$ is a result of an accumulated phase shift between two
extreme sites, as discussed below Eq.~(\ref{gen}).

\subsection{Scaling for symmetric states}
\label{abst_sym}

Our next step is to remove the optimization of the QFI over all
possible input states $|\psi\rangle$, and concentrate on a particular
family of states which are more relevant for current experimental
setups with ultracold atoms. 
We assume thus that  $N$ atoms are distributed among $M$ sites
symmetrically, i.e. the coefficient $C_{\vec n}$ is invariant upon
exchange of any two indices. A prominent member of such family of states is the
superfluid state \cite{rmp_zw}, where each atom is in a coherent superposition, i.e.
\begin{equation}\label{sf}
  |\psi_{\rm sf}\rangle=\frac{1}{\sqrt{N!M^{N}}}\left(\hat a^\dagger_1+\ldots+\hat a^\dagger_M\right)^{\otimes N}|0\rangle.
\end{equation}
This state can be written in the form (\ref{state}) with coefficients
\begin{equation}
  C_{\vec n}=\frac1{\sqrt{M^{N}}}\frac{\sqrt{N!}}{\sqrt{n_1!\ldots n_{M-1}!(N-n_1-\ldots -n_{M-1})!}},
\end{equation}
which clearly posses the aforementioned symmetry. Another example is a
symmetric Mott insulator state \cite{rmp_zw}, where each site is occupied by $N/M$ atoms. Below we calculate the sensitivity $\Delta^2\theta$
using a generic symmetric state without taking any specific value of $C_{\vec n}$'s.

To this end, we calculate the QFI using Eq.~(\ref{qfi}) and the
phase-transformation generator $\hat h$ for a linear perturbing
potential from Eq.~(\ref{gen}). The average is equal to
\begin{equation}
  \langle\psi|\hat h|\psi\rangle=\sum_{k=1}^Mk\langle\hat n_k\rangle=\frac12(M+1)N,\label{av}
\end{equation}
where in the last step, due to the imposed symmetry, we used $\langle\hat n_k\rangle=\frac NM$.
Calculation of the average of $\hat h^2$ involves a few intermediate
steps, which are presented in detail in Appendix \ref{app_QFI}. The
final result is that, for symmetric states and a linear perturbing potential, 
the QFI reads
\begin{equation}\label{fqvar}
  F_Q=\frac13(M+1)M^2\Delta^2\hat n,
\end{equation}
where $\Delta^2\hat n=\langle \hat n^2\rangle-\langle \hat n\rangle^2$ is the on-site variance of the atom number. 
The value of $\Delta^2\hat n$ can be easily calculated for some particular states. For instance,  
the uncorrelated super-fluid state (\ref{sf}) gives $\Delta^2\hat
n=\frac{M-1}{M^2}N$, and as a consequence the QFI scales at the SNL, i.e. $F_Q=\frac13(M^2-1)\cdot N$. 
A strongly entangled symmetric NOON state 
\begin{equation}
  |\psi_{\rm NOON}\rangle=\frac{|N,0,\ldots,0\rangle+|0,N,\ldots,0\rangle+\ldots+|0,\ldots,N\rangle}{\sqrt M}
\end{equation}
gives $\Delta^2\hat n=\frac{M-1}{M^2}N^2$, which gives the Heisenberg scaling of the QFI, $F_Q=\frac13(M^2-1)\cdot N^2$. 

These two examples suggest that for all symmetric states the 
on-site variance has a particularly simple form $\Delta^2\hat
n=\frac{M-1}{M^2}f(N)$, where $f(N)$ does not depend on $M$. Indeed, the above form of $\Delta^2\hat n$ can be explained as follows. 
When the ``volume'' of the system -- in this case the number of sites $M$ -- tends to infinity, the atom number fluctuations in one site tend to zero. 
On the other hand, when the whole system consists of only one site, the fluctuations are zero, since the number of atoms
in the system is fixed. The only possible dependence of the variance on $M$, which behaves correctly in both these regimes and also accounts for the scaling 
of the average value, $\langle\hat n\rangle^2=\frac{N^2}{M^2}$, is
thus $\Delta^2\hat n=\frac{M-1}{M^2}f(N)$. Putting this into Eq.~(\ref{fqvar}) gives
$F_Q=\frac13(M^2-1)f(N)$ and the sensitivity
\begin{equation}\label{fqm}
  \Delta^2\theta\geq\frac1m\frac1{f(N)}\frac3{M^2-1}.
\end{equation}
The function $f(N)$ can be interpreted as a measure
of the entanglement present in the system, related to the increased on-site atom number fluctuations. In the two-well case $M=2$, such usefully entangled
states which give $\Delta^2\hat n>N$ are called ``phase-squeezed'' \cite{grond}, 
because the growth of the number fluctuations is accompanied by a
decreased spread of the conjugate variable, which is the relative
phase between the wells. 

Most important, the scaling with $M$ of the sensitivity for
symmetric states (\ref{fqm}) is only slightly worse than the ultimate
scaling (\ref{opt}) obtained for the best possible state, and becomes
the same for large $M$.

\subsection{Scaling for non-linear potentials}
\label{nonlin}

In this part, we generalize the results of section \ref{abst_ult} for non-linear potentials $V(x)=g\,x^j$, 
where $j\in\mathbb{R}$. We assume that the first lattice site is placed at $x=x_0$ and define the interferometric phase
as $\theta=t\,g\,x_0^j/\hbar$. Accordingly, the phase imprinted on the $k$-th site is $k^j\theta$. From this, we can easily construct the generator as in Eq.~(\ref{gen}) which is
equal to
\begin{equation}
  \hat h_j=\sum_{k=1}^jk^j\hat n_k.
\end{equation}
The QFI can be again bounded by Eq.~(\ref{var}), giving 
\begin{equation}
  \Delta^2\theta\geq\frac1m\frac1{(\lambda_{\rm max}-\lambda_{\rm min})^2}. 
\end{equation}
The extreme eigenvalues for $j>0$ 
are equal to $\lambda_{\rm min}=N^2$ for $|\psi_{\rm min}\rangle$ and $\lambda_{\rm max}=M^jN^2$ for $|\psi_{\rm max}\rangle$ while for $j<0$ the minimal and maximal eigen-values are exchanged.
The states $|\psi_{\rm min}\rangle$ and $|\psi_{\rm max}\rangle$ are
still the ones defined in Eq.~(\ref{minmax}). The best possible
sensitivity is thus
\begin{equation}\label{theta_j}
  \Delta^2\theta\geq\frac1m\frac1{(M^j-1)^2}\frac1{N^2}. 
\end{equation}

\begin{figure}[htb!]
  \includegraphics[clip, scale=0.35]{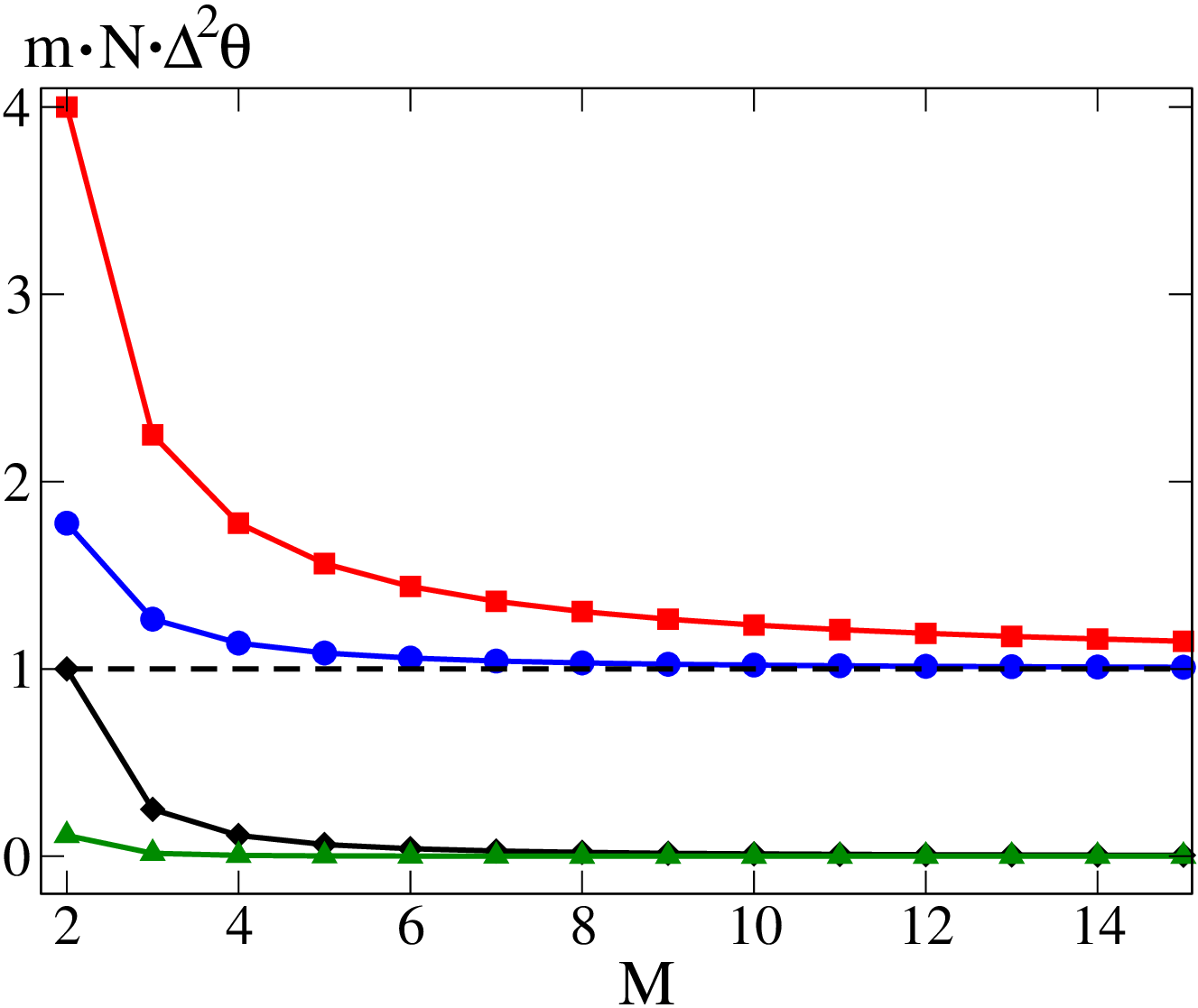}
  \caption{
   Ultimate scaling of the sensitivity for various non-linear potentials given by Eq.~(\ref{theta_j}). Red squares, blue circles, black diamonds
   and green triangles correspond to $j=-2,-1,1,2$,
   respectively. The black dashed line denotes the sensitivity of a standard $j=1,M=2$ two-mode interferometer.
  }\label{fig_nonlin}
\end{figure}
The behavior with $M$ of the r.h.s. of the above equation for various $j$ is show in
Fig.~\ref{fig_nonlin}.
When $j>0$, $\Delta^2\theta$ continuously improves with increasing
$M$, and the larger $j$ is, the quicker is the improvement. This would be the case -- for instance -- for the quadratic potential, i.e. $j=2$. On the other hand, when the potential 
decays with distance $j<0$, as is the case of the Casimir-Polder-type interactions which can exhibit the $j=-3$ or $j=-4$ scaling \cite{antezza,chwed2}, there is no substantial
gain from the multi-mode structure of the interferometer. This happens because already the phase imprinted on the second well, which is equal to $2^{-|j|}\theta$, is negligible when
compared to $\theta$. In this case
$\Delta^2\theta\geq\frac1m\frac1{N^2}$ and by increasing $M$ the sensitivity saturates to the value corresponding to
a linear potential and two wells, since effectively only the first well experiences the action of the
perturbing potential.

Apart from the ultimate scaling, it is interesting to derive an expression for the sensitivity for symmetric states, as was done in Section \ref{abst_sym} for linear
potentials. The derivation does not change substantially with respect to the line of reasoning presented in Appendix \ref{app_QFI}, with
the only difference that the summations over the sites involve $k^j$
instead of $k$. The resulting sensitivity bound for symmetric states reads
\begin{equation}\label{fqm_nonlin}
  \Delta^2\theta\geq\frac1m\frac1{f(N)}\frac{M^2}{4\left(M\rm{H}_{M}^{(-2j)}-(\rm{H}_{M}^{(-j)})^2\right)},
\end{equation}
where $\rm{H}_{n}^{(r)}=\sum_{k=1}^{n}k^{-r}$ is known as the Harmonic number.
For $j>0$ and a large number of lattice sites, the sensitivity bound
can be approximated by
\begin{equation}\label{fqm_nonlin_app}
  \Delta^2\theta\geq\frac1m\frac1{f(N)}\frac1{M^{2j}}\frac{(1+j)^2(1+2j)}{(2j)^2},
\end{equation}
which shows that the $1/M^j$ improvement on $\Delta\theta$ is
preserved for symmetric states.

We underline that -- via Eq.~(\ref{theta_j}) -- the multi-mode interferometer could serve as a sensitive tool to determine the spatial dependence of the perturbing potential, 
which is a relevant task for instance in the determination of the Casimir-Polder force or deviations from the Newton law of gravity at the micrometer scale.

\section{Sensitivity for the fit to the interference pattern}
\label{fit}

The QFI was used above to set the lower limit for the
sensitivity by optimizing of $\Delta^2\theta$ with respect to the output
measurement and estimator. 
Below we demonstrate that the scaling of the sensitivity with $M$,
which is present in Eq.~(\ref{fqm}), is preserved even for a
particularly simple phase estimation protocol, although the sensitivity
does not saturate the QFI. Also the scaling with $N$ is preserved for
the superfluid and phase-squeezed states.

We assume that after the system acquires the phase as in Eq.~(\ref{ev_op}), 
the optical lattice is switched off and the wave-packets $\psi_k(x)$ -- which were initially localized at each site -- 
expand and in the far field form an interference pattern. Positions of atoms are recorded and the phase is estimated by fitting the density 
$\rho(x|\theta)=\langle\hat\Psi^\dagger(x|\theta)\hat\Psi(x|\theta)\rangle$ to the acquired data. 
In our past work \cite{chwed_njp} we have shown that when this estimation protocol is employed, the sensitivity is given by
\begin{equation}
  \Delta^2\theta=\frac1m\frac{F_1+C}{F_1^2},\label{phil}
\end{equation}
where $F_1$ is the one-body Fisher information which reads
\begin{equation}
\label{fi}
  F_1=\int dx\frac1{\rho(x|\theta)}\left(\frac{\partial\rho(x|\theta)}{\partial\theta}\right)^2.
\end{equation}
The coefficient $C$ depends both on the one- and two-body correlations,
\begin{equation}
  C=\!\int\!\! dx\!\int\!\! dy\, G^{(2)}(x,y|\theta)\,\partial_\theta\log\left[\rho(x|\theta)\right]\,\partial_\theta\log\left[\rho(y|\theta)\right],
\end{equation}
where $G^{(2)}(x,y|\theta)=\langle\Psi^\dagger(x|\theta)\hat\Psi^\dagger(y|\theta)\Psi(y|\theta)\hat\Psi(x|\theta)\rangle$.
Our goal is to analyze how the sensitivity (\ref{phil}) depends on both the number of sites $M$ and the non-classical correlations present in the state (\ref{state}). 

In section \ref{abst_sym} we have seen that the usefully entangled states for the
present interferometer have enhanced number fluctuations. In
particular, the NOON state provides the best scaling with the number
of particles $N$. This statement, however, assumes that we are always
able to perform the optimal measurement, which in general depends
itself on the input state. In this section instead we fix the
measurement (as described above), and study the behavior of the sensitivity
for different states characterized by enhanced number fluctuations (we
will see that such states are still the useful ones also for the
particular measurement considered). 
For two lattice sites  ($M=2$), one can easily generate a
family of such usefully entangled states by numerically finding the ground state of the two-well Bose-Hubbard Hamiltonian
\begin{equation}
  \hat H_{\rm BH}=-E_J\hat J_x+U\hat J_z^2\label{bh}
\end{equation}
with $U<0$. Here, $E_J$ is the Josephson energy, related to the
frequency of the inter-well tunneling, and $U$ is the amplitude of the
on-site two-body interactions. This procedure for generating the input
states is also relevant for the setup we consider, since it will
correspond to adiabatical preparation in the lattice.
To perform an analogous calculation for higher $M$ is very demanding, since the dimensionality of the Hilbert space grows rapidly,
as indicated by multiple sums present in Eq.~(\ref{gen1}). 
This difficulty can be overcome by replacing the true state (\ref{gen1}) with 
\begin{equation}\label{prod}
  |\psi\rangle=\bigotimes_{k=1}^M\sum_{n_k=0}^\infty C_{n_k}|n_k\rangle.
\end{equation}
This is the Gutzwiller approximation of the quantum state of the bosons (see \cite{rmp_zw} and references therein). The number of atoms is not fixed anymore, and thus
each site is independent from the others. 
Although such state does not recover correlations among the wells, it can be used to mimic non-classical correlations by increasing the {\it on-site} atom number variance $\Delta^2\hat n$.
In analogy to the fixed-$N$ case (\ref{state}) we assume that the average number of atoms per site is $\langle\hat n_k\rangle=\frac NM$, so that the average number of atoms in the
whole system is $N$. 


We express the field operator $\hat\Psi(x|\theta)$ in the far field, 
\begin{equation}\label{op_ff}
  \hat\Psi(x|\theta)=\tilde\psi\left(\frac x{\tilde\sigma^2}\right)\sum_{k=1}^Me^{-i\phi(x)\cdot k}\hat a_k.
\end{equation}
We assumed that initially all the wave-packets $\psi_k(x)$ had identical shape but were localized around separated sites of the optical lattice. The function 
$\tilde\psi$ is the Fourier transform of $\psi_k(x)$, which is common for each well, 
and $\tilde\sigma=\sqrt{\frac{\hbar t}{\mu}}$ ($\mu$ is the atomic mass and $t$ is the expansion time). The phase factor is defined as
\begin{equation}\label{phi}
  \phi(x)=2\frac{x_0}{\tilde\sigma^2}x+\theta
\end{equation}
where $x_0$ is the lattice spacing first introduced below Eq.~(\ref{ham}). 

The density $\rho(x|\theta)$ obtained using Equations (\ref{prod}) and (\ref{op_ff}) is
\begin{equation}
  \rho(x|\theta)=\langle\psi|\hat\Psi^\dagger(x|\theta)\hat\Psi(x|\theta)|\psi\rangle=N\Big|\tilde\psi\left(\frac x{\tilde\sigma^2}\right)\Big|^2\cdot f(x|\theta)\label{p1},
\end{equation}
where $|\tilde\psi|^2$ is the envelope of the interference pattern, while the fringes are contained in
\begin{equation}\label{f}
  f(x|\theta)=1+\frac{\langle\hat a\rangle^2}{N}\left(\frac{\sin^2\left(\frac{M\phi(x)}2\right)}{\sin^2\left(\frac{\phi(x)}2\right)}-M\right).
\end{equation}
The density is put into Eq.~(\ref{fi}), giving
\begin{equation}
  F_1=N\int dx \Big|\tilde\psi\left(\frac x{\tilde\sigma^2}\right)\Big|^2
  \frac1{f(x|\theta)}\left(\frac{\partial f(x|\theta)}{\partial\theta}\right)^2.
\end{equation}
When the density of the interference pattern consists of many fringes, the envelope $|\tilde\psi|^2$ varies much slower in $x$ then the function $f(x|\theta)$. 
An integral of a slowly varying envelope with a quickly oscillating periodic function can be split into a product of 
an integral of the envelope alone times the average value of the quickly changing $f(x|\theta)$. Since the envelope is normalized, we immediately get
\begin{equation}
  F_1=N\int_0^{2\pi}\frac{d\phi}{2\pi}\frac1{f(\phi)}\left(\frac{\partial f(\phi)}{\partial\phi}\right)^2,\label{f1}
\end{equation}
where $\phi$ was defined in Eq.~(\ref{phi}). 
This integral depends on the quantum state through the average value $\langle\hat a\rangle$, as can be seen from Eq.~(\ref{p1}). Apart from the analytical case $M=2$, its value
has to be calculated numerically. 

Now we have to calculate the coefficient $C$, which in turn requires to
compute the second-order correlation function. After a lengthy calculation we obtain (see Appendix \ref{APP_C} for details)
\begin{subequations}\label{C}
  \begin{eqnarray}
    \!\!\!\!\!\!\!\!\!C&=&\left[\langle\hat a^2\rangle^2\!-\!\left(\frac NM\right)^2\!+\!2\frac NM\langle\hat a\rangle^2\!-\!2\langle\hat a^2\rangle\langle\hat a\rangle^2\right]I_1\\
    &+&\left[2\langle\hat a^2\rangle\langle\hat a\rangle^2-2\frac NM\langle\hat a\rangle^2\right]I_2.
  \end{eqnarray}
\end{subequations}
where the two integrals are given by
\begin{subequations}\label{i1}
  \begin{eqnarray}
  I_1&=&\int_{-\pi}^\pi\frac{d\phi}{2\pi}\int_{-\pi}^\pi\frac{d\phi'}{2\pi}\ \partial_\theta\log \left[f(\phi)\right]\partial_\theta \left[f(\phi')\right]\\
  &\times&\frac{\cos[M(\phi+\phi')]-1}{\cos[\phi+\phi']-1}
  \end{eqnarray}
\end{subequations}
and
\begin{subequations}\label{i2}
  \begin{eqnarray}
  I_2&=&\int_{-\pi}^\pi\frac{d\phi}{2\pi}\int_{-\pi}^\pi\frac{d\phi'}{2\pi}\ \partial_\theta\log \left[f(\phi)\right]\partial_\theta \left[f(\phi')\right]\\
  &\times&\frac{\sin\left[\frac M2\phi\right]\sin\left[\frac M2\phi'\right]\sin\left[\frac M2(\phi+\phi')\right]}
  {\sin\left[\frac\phi2\right]\sin\left[\frac{\phi'}2\right]\sin\left[\frac{\phi+\phi'}2\right]}.
  \end{eqnarray}
\end{subequations}

As a first test of Equations (\ref{f1}) and (\ref{C}), in the next
section we show that for $M=2$ we recover a known analytical expression for the sensitivity obtained using a two-mode state with fixed $N$
\cite{chwed_njp}. 

\subsection{Fit sensitivity with two lattice sites}

To make a direct comparison, we start by recalling the main result of our previous work \cite{chwed_njp}, where we focused on phase estimation with a fixed-$N$ two-mode state 
$|\psi\rangle=\sum_{n=0}^NC_n|n,N-n\rangle$. In the resulting $N$
qbits space, every possible interferometric operation can be mapped
into a rotation, with the angular momentum operators
$\hat J_x=\frac12(\hat a^\dagger\hat b+\hat a\hat b^\dagger)$, $\hat J_y=\frac1{2i}(\hat a^\dagger\hat b-\hat a\hat b^\dagger)$ and $\hat J_z=\frac12(\hat a^\dagger\hat a-\hat b^\dagger\hat b)$. 
We have shown that when the phase is deduced from the fit of the one-body density to the interference pattern
formed by two expanding wave-packets, the sensitivity (\ref{phil}) can be expressed in terms of two quantities. One 
is the spin-squeezing parameter \cite{grond}, which is a measure of entanglement related to the phase-squeezing
of $|\psi\rangle$ and reads $\xi_\phi=\sqrt{N\frac{\Delta^2\hat J_y}{\langle\hat J_x\rangle^2}}$. The other is the 
visibility of the interference fringes, $\nu=\frac2N\langle\hat J_x\rangle$. The sensitivity 
\begin{equation}\label{fin}
  \Delta^2\theta=\frac1{mN}\left[\xi_\phi^2+\frac{\sqrt{1-\nu^2}}{\nu^2}\right]
\end{equation}
results from the interplay between the growing entanglement ($\xi^2_\phi\rightarrow0$) and the loss of the visibility ($\nu\rightarrow0$). Naturally, there is some optimally squeezed state, 
which gives the minimal value of $\Delta^2\theta$, as discussed in detail in \cite{chwed_njp}.

\begin{figure}[htb!]
  \includegraphics[clip, scale=0.33]{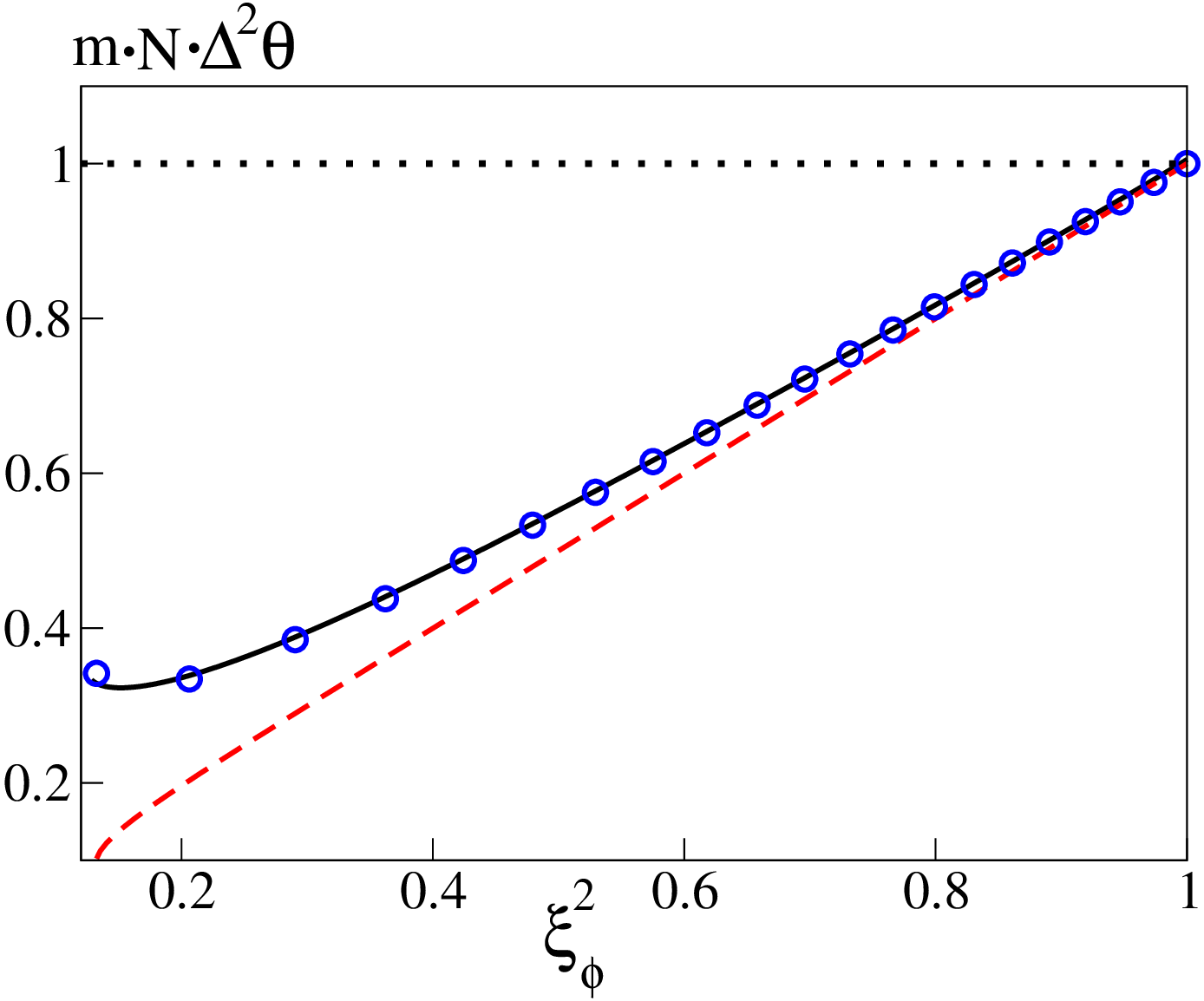}
  \caption{
    The sensitivity $m\cdot N\cdot\Delta^2\theta$ of the parameter estimation from the fit to the density of the interference pattern with $M=2$ and $N=200$ 
    calculated using the state with fixed-
    (open blue circles) and non-fixed number of atoms (solid black line) as a function of the phase-squeezing parameter. The sensitivities shown are normalized to the SNL. 
    The dashed red line shows the ultimate precision of the
    phase estimation given by the QFI from Eq.~(\ref{fqvar}). The horizontal black dotted line shows the shot noise limit. 
  }\label{fig_comp}
\end{figure}

We now calculate the sensitivity (\ref{phil}) using the Gutzwiller
approximation (\ref{prod}) for two wells, and check how it compares
with the exact result (\ref{fin}). 
As shown in the Appendix \ref{app_two}, in this case, both the one body Fisher information (\ref{f1}) and the coefficient $C$ (\ref{C})
can be evaluated analytically and the outcome is
\begin{equation}\label{finprod}
  \Delta^2\theta=\frac1{mN}\left[\tilde\xi_\phi^2+\frac{\sqrt{1-\tilde\nu^2}}{\tilde\nu^2}\right],
\end{equation}
where the tilde sign denotes the phase-squeezing parameter and the visibility calculated with a product state. Although Equations (\ref{fin}) and (\ref{finprod}) have identical form,
we still have to verify whether the two-well entanglement of the
fixed-$N$ state can be reproduced with a product state by increasing the on-site fluctuations. 
We model the coefficients of Eq.~(\ref{prod})  with a Gaussian
\begin{equation}\label{gauss}
  C_{n_k}\propto e^{-\frac{\left(n_k-\langle n\rangle\right)^2}{2\sigma^2}},
\end{equation}
where $\langle n\rangle=\frac N2$ for a two-mode state.
The on-site atom number variance is simply $\Delta^2\hat n=\sigma^2$, thus a coherent superfluid state can be modeled by setting the fluctuations at the Poissonian level, i.e. 
$\sigma=\sqrt{N}$. To check if Eq.~(\ref{finprod})
will drop below the SNL when the fluctuations are super-Poissonian, we proceed as follows. For various values of $\sigma>\sqrt{N}$ we calculate
the phase-squeezing parameter $\tilde\xi_\phi^2$ and the visibility $\tilde\nu$ and plot the sensitivity (\ref{finprod}) as a function of $\tilde\xi_\phi^2$. To compare with
the fixed-$N$ case, we find the ground state of the Hamiltonian (\ref{bh}) for various values of the ratio $\frac{E_J}U<0$. For every state found in this way, we calculate
the phase-squeezing parameter $\xi_{\phi}^2$ and the visibility $\nu$ and on the same plot draw the sensitivity (\ref{fin}) as a function of $\xi_\phi^2$. 
Figure \ref{fig_comp} shows that not only Equations (\ref{finprod}) and (\ref{fin}) have identical form but also the quantum correlations giving sub shot-noise sensitivity can be
perfectly mimicked using a product state (\ref{prod}) with large
on-site atom number fluctuations. In both cases, the sensitivity has
an optimal point, where the entanglement is balanced by the fringe
visibility. Clearly, the Gutzwiller approximation gets worse when
the correlations between different wells increase (that is, going
further to the left in Fig.~\ref{fig_comp}), but for the usefully
states we consider and for this particular choice of the measurement, it is exact for all practical purposes. 

Having proven its usefulness, in the following section we employ the
Gutzwiller ansatz (\ref{prod}) plus the
Gaussian approximation on the coefficient of the product state
(\ref{gauss}) in order to
analyze how the sensitivity (\ref{phil}) behaves for higher $M$, where results from a fixed-$N$ model are unavailable.

\subsection{Fit sensitivity with $M$ lattice sites}

\begin{figure}[htb!]
  \includegraphics[clip, scale=0.33]{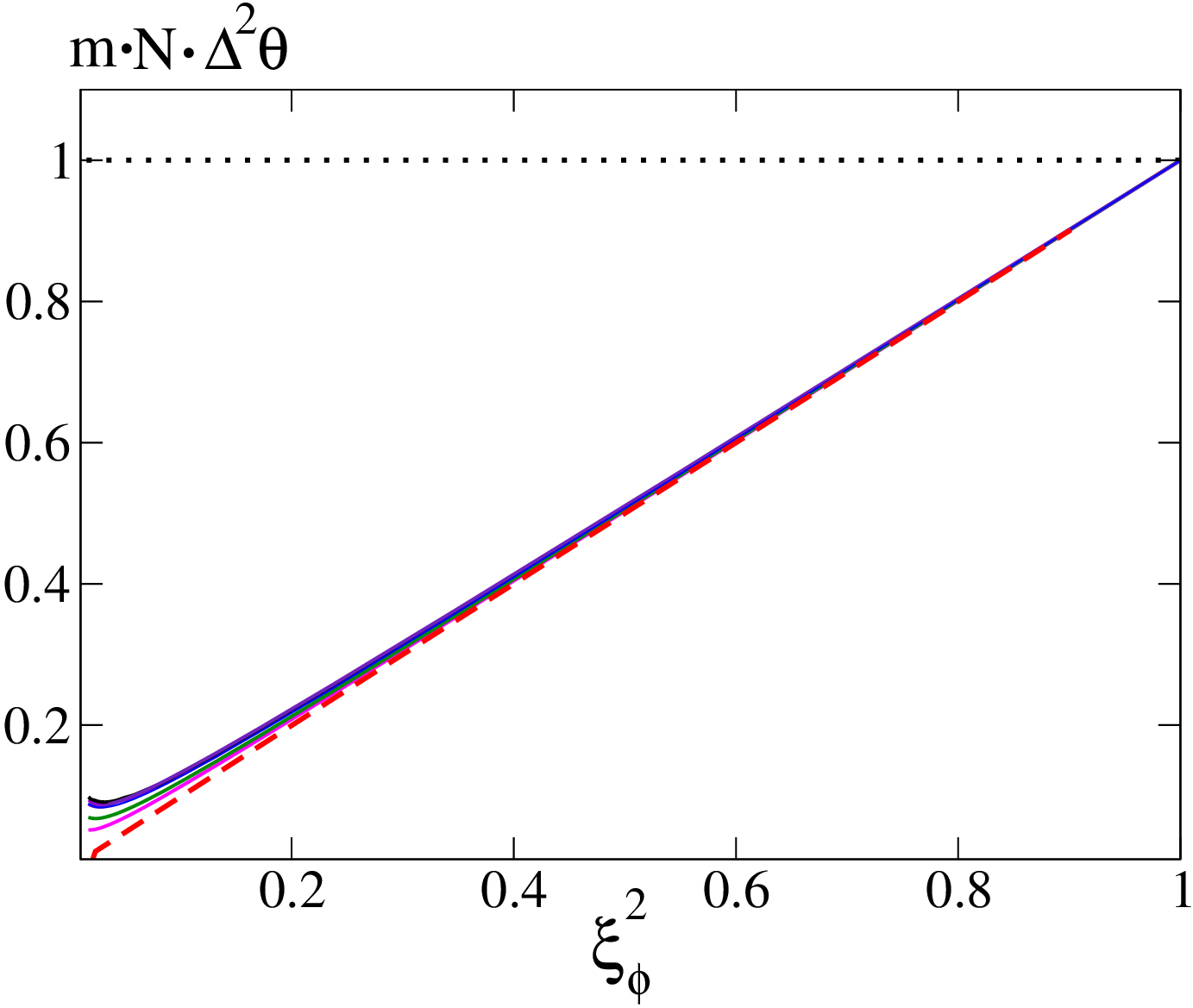}
  \caption{
    (color online) The sensitivity 
    $m\cdot N\cdot\Delta^2\theta$ of the parameter estimation from the fit to the density of the interference pattern for $M=2,4,6,8,10$ and 12 (different color solid lines)
    with a fixed average number of atoms equal to $\langle N\rangle=5\times10^4$. The sensitivities shown are normalized to the SNL and multiplied by the factor $\frac13(M^2-1)$. 
    The dashed red line shows the ultimate precision of the
    phase estimation given by the QFI from Eq.~(\ref{fqvar}). The horizontal black dotted line shows the shot noise limit. 
  }\label{fig_M}
\end{figure}

For $M>2$, when the integrals (\ref{f1}),  (\ref{i1}) and (\ref{i2}) cannot be evaluated analytically, we calculate the sensitivity (\ref{phil}) numerically. 
We fix the average total number of atoms to be $\langle\hat N\rangle=5\times10^{4}$ and
for a given $M$ we calculate the sensitivity (\ref{phil}) for different values of the on-site atom-number fluctuations using the Gaussian model (\ref{gauss})
with $\langle n\rangle=\frac{\langle \hat N\rangle}M$. 
In order to compare the results
for different $M$, we normalize the sensitivity by the factor
$\frac13(M^2-1)$, that is, by the sensitivity achieved with the best
possible measurement (\ref{fqm}) at the shot-noise level. In
Fig.~\ref{fig_M} we plot $m\langle\hat N\rangle\times\frac13(M^2-1)\Delta^2\theta$ for $M=2,4,6,8,10$ and 12.
The outcome is -- with a high level of accuracy -- independent of $M$. We thus conclude that
the sensitivity of the fit has a form similar to (\ref{fqm}), i.e.
\begin{equation}
  \Delta^2\theta=\frac1m\frac1{\zeta(\langle\hat N\rangle)}\frac3{(M^2-1)},
\end{equation}
where $\zeta(\langle\hat N\rangle)$ is some function of the average
total number of atoms, which becomes larger for states with increased on-site atom
number fluctuations,
$\zeta>\langle\hat N\rangle$.

In Fig.~\ref{fig_M}, as the phase-squeezing parameter drops, the
discrepancy between the curves calculated for different $M$
grows. This is a spurious result due to the Gutzwiller approximation (\ref{prod}) for the state $|\psi\rangle$. When the average number of atoms in the system is fixed and $M$ is changed, the average on-site occupation varies.
This is not a problem for as long as the width $\sigma$ of the Gaussian (\ref{gauss}) is small as compared to $\langle\hat n\rangle$. However, we increase $\sigma$ to mimic 
the growing entanglement in the system. Once $\sigma$ is comparable to
$\langle\hat n\rangle$, the change in the on-site population for
different $M$ becomes relevant. Therefore, by changing $M$ we are
actually changing the nature of the quantum state and, in particular, we cannot 
assume that the squeezing parameter remains fixed.
One could avoid this problem by -- rather than fixing $\langle N\rangle$ -- fixing the on-site population $\langle n\rangle$ and increase the number of wells so that 
$\langle N\rangle=M\langle n\rangle$.

\section{Conclusions}

We have analyzed the performance of  ultracold bosons trapped
inside an optical lattice as an interferometric device for the
measurement of spatially dependent forces. For a
potential $g\,x^{j}$ with $j>0$, we showed that, optimizing over all
possible output measurements,
the coupling constant $g$ can be measured with an
uncertainty $\Delta g$ scaling like $M^{-j}$, for a large number of
lattice sites $M$, and independent of the quantum state employed.
We then showed that this scaling with the number of lattice sites can
still be reached with a particular output measurement which is
commonly employed with ultracold atoms: a least square fit to the
interference pattern resulting from the overlap o the freely expanding
atoms released from the confining potential. 
We also derived the improvements coming from entangling the quantum
state.

The device here studied is an example of a matter-wave multi-path
interferometer. While the sensitivity bounds for a general two-path
interferometer are already largely studied, the same analysis based
on rigorous estimation theory results has not yet been performed for
multi-path devices.
Moreover, the potential of the implementation of these devices with
atoms seems very promising, due to the growing control and tunability
of the available trapping and probing tools, especially for ultracold
atoms.

Finally, we underline that the precision of the time-of-flight interferometer, where the phase is estimated from the fit of the averaged density to the measured interference pattern 
cannot be surpassed by adding a hopping term to the generating Hamiltonian (\ref{ham}). The presence of the hopping -- which induces the Bloch oscillations --
is an extra complication, which changes the optimal states depending on
the ratio between the tunneling energy to the phase related energy $\epsilon$, but does not introduce any further benefit. Some of these conclusions can be found in \cite{chwed2} and
will be discussed in detail elsewhere.



\section{Acknowledgments}
We thank A. Recati for some very useful remarks.
J. Ch. acknowledges Foundation for Polish Science International TEAM Program co-financed by the EU European Regional Development Fund and the support of the National Science Centre.
F.P. acknowledges support by the Alexander Von Humboldt foundation.
A.S. acknowledges support of the EU-STREP Project QIBEC.
QSTAR is the MPQ, IIT, LENS, UniFi joint center for Quantum Science and Technology in Arcetri. 

\appendix

\section{Calculation of the QFI}
\label{app_QFI}

In order to calculate the $F_Q$, we need to derive an expression for the variance of the operator $\hat h$. Using the definition (\ref{gen}), we get
\begin{equation}
  \langle\hat h^2\rangle=\sum_{k_1,k_2}k_1k_2\langle\hat n_{k_1}\hat n_{k_2}\rangle=
  \sum_{k}k^2\langle\hat n_k^2\rangle+\sum_{k_1\neq k_2}k_1k_2\langle\hat n_{k_1}\hat n_{k_2}\rangle.
\end{equation}
We now make of the symmetry between the wells. The average value $\langle\hat n_k^2\rangle$ does not depend on index $k$ and
thus can be denoted as $\langle\hat n^2\rangle$. The same argument concerns the inter-well correlation $\langle\hat n_{k_1}\hat n_{k_2}\rangle$, which will be denoted as
$\langle\hat n_i\hat n_j\rangle$, where $i,j$ are indeces of {\it any} two sites. The above expression thus boils down to
\begin{eqnarray}
  &&\langle\hat h^2\rangle=\langle n^2\rangle\sum_{k}k^2+\langle\hat n_i\hat n_j\rangle\sum_{k_1\neq k_2}k_1k_2=\\
  &&\frac M6(M+1)(2M+1)\langle\hat n^2\rangle+\langle\hat n_i\hat n_j\rangle\frac M{12}(M^2-1)(3M+2).\nonumber
\end{eqnarray}
Our next step is to show that  $\langle\hat n_i\hat n_j\rangle$ can be written as a function of $\langle\hat n^2\rangle$ only. To this end, since the state is symmetric, 
we can take $i=1$ and $j=M$ and obtain
\begin{eqnarray}
  \langle\hat n_1\hat n_M\rangle&=&\sum_{\vec n}C_{\vec n}^2\,n_1\cdot(N-n_1-\ldots -n_{M-1})\\
  &=&\frac{N^2}M-\sum_{\vec n}C_{\vec n}^2\,n_1\cdot(n_1+\ldots+n_{M-1}).
\end{eqnarray}
Now, since it doesn't matter which two wells we correlate, in the sum instead of $n_1$ we can use an average over $M-1$ wells as follows
\begin{equation}
  \langle\hat n_1\hat n_2\rangle=\frac{N^2}M-\sum_{\vec n}C_{\vec n}^2\,\frac{n_1+\ldots+n_{M-1}}{M-1}\cdot(n_1+\ldots+n_{M-1}).
\end{equation}
Finally, we exploit the fact that the total number of atoms is fixed and thus $n_1+\ldots+n_{M-1}=N-n_M$ and obtain
\begin{eqnarray}
  \langle\hat n_1\hat n_M\rangle&=&\frac{N^2}M-\sum_{\vec n}C_{\vec n}^2\,\frac{(N-n_M)^2}{M-1}\\
  &=&\frac{N^2}M-\frac{\langle n^2\rangle-2N^2/M+N^2}{M-1}.
\end{eqnarray}
Using the result for the average value of $\hat h$, we finally obtain
\begin{equation}
  F_Q=4\Delta^2\hat h=\frac13(M+1)M^2\Delta^2\hat n,
\end{equation}
where $\Delta^2\hat n=\langle\hat n^2\rangle-\langle\hat n\rangle^2$ is the on-site variance.

\section{Calculation of the sensitivity}
\label{APP_C}

In this Appendix we give details of derivation of coefficient $C$ from Eq.~(\ref{C}) which is used to evaluate the sensitivity (\ref{phil}) using the product state (\ref{prod}).
In order to calculate $C$, one first has to find the second order correlation function,
\begin{equation}
  G^{(2)}(x,y|\theta)=\langle\hat\Psi^\dagger(x|\theta)\hat\Psi^\dagger(y|\theta)\hat\Psi(y|\theta)\hat\Psi(x|\theta)\rangle.
\end{equation}
When the field operator, which is introduced in Eq.~(\ref{op_ff}), is put into this definition, we obtain
\begin{eqnarray}
  &&G^{(2)}(x,y|\theta)=\Big|\tilde\psi\left(\frac x{\tilde\sigma^2}\right)\Big|^2\Big|\tilde\psi\left(\frac y{\tilde\sigma^2}\right)\Big|^2\times\\
  &&\!\!\!\sum_{k_1,k_2,k_3,k_4=1}^M\!\!\!\!\!\!e^{i\phi(x)(k_1-k_4)}e^{i\phi(y)( k_2-k_3)}\langle\hat a^\dagger_{k_1}\hat a^\dagger_{k_2}\hat a_{k_3}\hat a_{k_4}\rangle,
\end{eqnarray}
where $\phi(x)=2\frac{x_0}{\tilde\sigma^2}x+\theta$.
The above sum cannot be performed in a straightforward manner, since special care has to be taken to distinguish cases when the indices $k_1\ldots k_4$ in the average value
$\langle\hat a^\dagger_{k_1}\hat a^\dagger_{k_2}\hat a_{k_3}\hat a_{k_4}\rangle$
are equal or different from each other. A careful classification of all the possible terms gives
\begin{widetext}
  \begin{subequations}\label{g2}
    \begin{eqnarray}
      &&G^{(2)}(x,y|\theta)=\Big|\tilde\psi\left(\frac x{\tilde\sigma^2}\right)\Big|^2\Big|\tilde\psi\left(\frac y{\tilde\sigma^2}\right)\Big|^2\times
      \left\{M\langle\hat a^\dagger\hat a^\dagger\hat a\hat a\rangle+(f(x+y)-M)\langle\hat a^\dagger\hat a^\dagger\rangle^2+\langle\hat n\rangle^2(f(x-y)+M^2)\right.\\
        &&+2\left[f(x)+f(y)-2M\right]\langle\hat a^\dagger\hat a^\dagger \hat a\rangle  \langle\hat a\rangle+
        \langle\hat n\rangle\langle\hat a\rangle^2\left[8M-2M^2+(M-4)\left(f(x)+f(y)\right)-2f(x-y)+2g(x,-y)\right]\\
      &&+\langle\hat a\rangle^2\langle\hat a^2\rangle\left[4M-2f(x)-2f(y)-2f(x+y)+2g(x,y)\right]\\
        &&\left.+\langle\hat a\rangle^4\left[f(x)f(y)-(M-4)(f(x)+f(y))+f(x+y)+f(x-y)-2g(x,y)-2g(x,-y)+M^2-6M\right]\right\}.
    \end{eqnarray}
  \end{subequations}
\end{widetext}
The function $f(x)$ is defined as
\begin{equation}
  f(x)=\frac{\cos\left[M\phi(x)\right]-1}{\cos\left[\phi(x)\right]-1},
\end{equation}
while the $g(x,y)$ function is
\begin{equation}
  g(x,y)=\frac{\sin\left[\frac{M\phi(x)}2\right]\sin\left[\frac{M\phi(y)}2\right]\sin\left[\frac{M(\phi(x)+\phi(y))}2\right]}
  {\sin\left[\frac{\phi(x)}2\right]\sin\left[\frac{\phi(y)}2\right]\sin\left[\frac{\phi(x)+\phi(y)}2\right]}.
\end{equation}
The coefficient $C$ is defined as a double integral of the above second order correlation function with the one-body probabilities calculated at positions $x$ and $y$, namely
\begin{equation}\label{coeff}
  C=\!\int\!\! dx\!\int\!\! dy\, G^{(2)}(x,y|\theta)\,\partial_\theta\log\left[\rho(x|\theta)\right]\,\partial_\theta\log\left[\rho(y|\theta)\right].
\end{equation}
The one-body density is defined in Eq.~(\ref{p1}). We use the fact that the interference pattern consists of many fringes, so that the envelope $|\tilde\psi|^2$ varies slowly
as compared to the oscillatory functions present in $f$, $g$ and the density $\rho$. Therefore, the integral (\ref{coeff}) can be rewritten as
\begin{subequations}
  \begin{eqnarray}
    C&=&\!\int_{-\pi}^\pi\frac{d\phi}{2\pi}\int_{-\pi}^\pi\frac{d\phi'}{2\pi}\, G^{(2)}(\phi,\phi'|\theta)\\
    &\times&\partial_\theta\log\left[\rho(\phi|\theta)\right]\,\partial_\theta\log\left[\rho(\phi'|\theta)\right],
  \end{eqnarray}
\end{subequations}
where we changed the variables $x\rightarrow\phi$ and $y\rightarrow\phi'$. It can be demonstrated that 
\begin{subequations}
  \begin{eqnarray}
    \int_{-\pi}^\pi\frac{d\phi}{2\pi}\partial_\theta\log\left[\rho(\phi|\theta)\right]=0,
  \end{eqnarray}
\end{subequations}
therefore all the terms from Equations (\ref{g2}) which do not depend on both $\phi$ and $\phi'$ are zero. Using a property of the functions $f$ and $g$
\begin{eqnarray*}
  &&\int_{-\pi}^\pi\frac{d\phi}{2\pi}\int_{-\pi}^\pi\frac{d\phi'}{2\pi}f(\phi+\phi')\partial_\theta\log\left[\rho(\phi|\theta)\right]\,\partial_\theta\log\left[\rho(\phi'|\theta)\right]=\\
  &&-\int_{-\pi}^\pi\frac{d\phi}{2\pi}\int_{-\pi}^\pi\frac{d\phi'}{2\pi}f(\phi-\phi')\partial_\theta\log\left[\rho(\phi|\theta)\right]\,\partial_\theta\log\left[\rho(\phi'|\theta)\right]
\end{eqnarray*}
and the same with $g(\phi,\phi')$ and $g(\phi,-\phi')$ we finally get
\begin{subequations}
  \begin{eqnarray}
    \!\!\!\!\!C&=&\left[\langle\hat a^2\rangle^2-\langle\hat n\rangle^2+2\langle\hat n\rangle\langle\hat a\rangle^2-2\langle\hat a^2\rangle\langle\hat a\rangle^2\right]\times\\
    &\times&\int_{-\pi}^\pi\frac{d\phi}{2\pi}\int_{-\pi}^\pi\frac{d\phi'}{2\pi}\frac{p'(\phi)}{p(\phi)}\frac{p'(\phi')}{p(\phi')}f(\phi+\phi')\\
    &+&\left[2\langle\hat a^2\rangle\langle\hat a\rangle^2-N\langle\hat a\rangle^2\right]\times\\
    &\times&\int_{-\pi}^\pi\frac{d\phi}{2\pi}\int_{-\pi}^\pi\frac{d\phi'}{2\pi}\frac{p'(\phi)}{p(\phi)}\frac{p'(\phi')}{p(\phi')}g(\phi,\phi'),
  \end{eqnarray}
\end{subequations}
which coincides with Eq.~(\ref{C}).

\section{Sensitivity for two wells}
\label{app_two}

First, we calculate the one-body Fisher information defined in Eq.~(\ref{f1}). For $M=2$ the $f$ function from Eq.~(\ref{f}) reads
\begin{equation}\label{f_prod}
  f(\phi)=1+\tilde\nu\cos\phi,
\end{equation}
where $\tilde\nu$ is the fringe visibility calculated with a product state, 
\begin{equation}
  \tilde\nu=\frac 2N\langle\hat J_x\rangle=\frac 2N\frac{\langle\hat a^\dagger\rangle\langle\hat b\rangle+\langle\hat a\rangle\langle\hat b^\dagger\rangle}2=
  \frac 2N\langle\hat a\rangle^2,\label{vis}
\end{equation}
where we used the symmetry between the two wells, and the last step was possible because the coefficients $C_n$ are taken to be real. 
We insert (\ref{f_prod}) into (\ref{f1}) and obtain an analytical outcome
\begin{equation}\label{tildeF1}
  F_1=1-\sqrt{1-\tilde\nu^2}.
\end{equation}
The integrals $I_{1}$ from Eq.~(\ref{i1}) and $I_2$ from Eq.~(\ref{i2}) can be calculated in a similar way, giving
\begin{equation}
  I_1=I_2=\frac2{\tilde\nu^2}(-2+\tilde\nu^2+2\sqrt{1-\tilde\nu^2}).
\end{equation}
As a result the coefficient $C$ from Eq.~(\ref{C}) for $M=2$ reads
\begin{equation}\label{tildeC0}
  C=\frac2{\tilde\nu^2}(-2+\tilde\nu^2+2\sqrt{1-\tilde\nu^2})\left[\langle\hat a^2\rangle^2-\frac{N^2}4\right].
\end{equation}
Finally, we note that for a symmetric state $\langle\hat J_y\rangle=0$, thus the variance of the $\hat J_y$ operator calculated using a product state reads
\begin{equation}
  \Delta^2\hat J_y=\frac12\left[\frac{N^2}4+\frac N2-\langle\hat a^2\rangle^2\right],
\end{equation}
which allows to rewrite Eq.~(\ref{tildeC0}) in a following way
\begin{equation}\label{tildeC}
  C=\frac4{\tilde\nu^2}(-2+\tilde\nu^2+2\sqrt{1-\tilde\nu^2})\left[\frac N4-\tilde\Delta^2\hat J_y\right].
\end{equation}
Combining (\ref{tildeF1}) and (\ref{tildeC}) as in Eq.~(\ref{phil}) we obtain
\begin{equation}
  \Delta^2\theta=\frac1{mN}\left[\tilde\xi_\phi^2+\frac{\sqrt{1-\tilde\nu^2}}{\tilde\nu^2}\right],
\end{equation}
where $\tilde\xi_\phi^2$ is the phase-squeezing parameter calculated with a product state.

\end{document}